\documentclass[twocolumn,preprintnumbers,amsmath,amssymb,
showkeys,floatfix,aps]{revtex4}
\usepackage{graphicx}
\usepackage{dcolumn}
\usepackage{bm}
\usepackage{bigints}
\usepackage{natbib}
\usepackage{amsmath}
\usepackage{hyperref}
\begin{document}
{ \footnotesize \quad Phys. Rev. D 107, 063506 \quad Published 6 March 2023 \qquad \qquad \qquad \qquad \qquad \qquad \qquad \qquad \href{https://link.aps.org/doi/10.1103/PhysRevD.107.063506}{DOI: 10.1103/PhysRevD.107.063506}}

\title{Separation of  CMB  {\Large $\mu$} spectral distortions from foregrounds
with poorly defined spectral shapes}

\author{D.~I. Novikov}
\affiliation{Astro-Space Center of P.N. Lebedev Physical Institute, Profsoyusnaya 84/32, Moscow, Russia 117997}
\author{A.O. Mihalchenko}
\affiliation{Astro-Space Center of P.N. Lebedev Physical Institute, Profsoyusnaya 84/32, Moscow, Russia 117997}
\affiliation{Moscow Institute of Physics and Technology, Institutskiy pereulok, d.9, Dolgoprudny, Moscow Region, 141701, Russia}
\begin{abstract}

 This paper proposes a new approach to separate the $\mu$ spectral distortions
 of the cosmic microwave background from foregrounds with poorly defined
 spectral shapes. The idea is based on finding the optimal response to the
 observed signal. This response is weakly sensitive to foregrounds with
 parameters that are within some certain limits of their possible variations
 and, at the same time, very sensitive to the amplitude of $\mu$ distortion.
 The algorithm described in this paper is stable, easy to implement, and
 simultaneously minimizes the response to foregrounds and photon noise.

\end{abstract}


\maketitle

\section{Introduction}
The detection of distortions in the frequency
spectrum of the cosmic microwave background (CMB) radiation is one of the key
tasks of observational cosmology \citep{2021ExA....51.1515C,
  2012MNRAS.419.1294C,2014Sci...344..586S,2016JCAP...03..047D,
  2014PTEP.2014fB107T}.
Deviations of the CMB spectrum from the blackbody shape represent a completely
new channel of information about the
fundamental physical processes in the early Universe, sometimes inaccessible
to other observations \citep{1969Ap&SS...4..301Z,
1991A&A...246...49B,2019BAAS...51c.184C,2018PhRvD..97d3525N}.

The epoch of $\mu$ distortions \citep {1970Ap&SS...7...20S} in the Universe takes place
in the interval of redshifts
between $z=2\times 10^6$ and $z=10^5$. The detection of such distortions can
provide essential information about the mechanisms of a possible energy
injection into the plasma during this period of time
\citep{1970Ap&SS...9..368S,1991ApJ...371...14D,1994ApJ...430L...5H,
  2012MNRAS.425.1129C,2012MNRAS.419.1294C,2014JCAP...10..029O}. 
At this stage the total number of photons in the
Universe remains unchanged, and the energy exchange between electrons
and photons is described by the Kompaneets equation \citep{1957JETP...4.730K}.
Therefore, any energy release leads to heating of photons while maintaining
their total number, which means a deviation from blackbody distribution
in the form of the Bose-Einstein spectrum with a nonzero chemical potential
(or $\mu$ distortion). Proposed missions targeting spectral distortions
are described in \citep{2016SPIE.9904E..0WK,2021PhyU...64..386N}.

The task of measuring $\mu$ distortions is very challenging and complicated
by the presence of foregrounds of various origin \citep{2017MNRAS.471.1126A}.
The spectra created by some
foregrounds as well as by the optical system of the telescope are poorly
predictable. In reality, the observed cosmic foreground spectrum (even
for a single line of sight) is a superposition of spectra with different
parameters (for example, with different dust temperatures).  Such a “cocktail”
of spectra is difficult or even impossible to estimate and predict with the
accuracy required for $\mu$ distortion measurements  \citep{2016MNRAS.460..227C, 2017MNRAS.471.1126A, 2019PhRvD.100j3508M, 2014PhRvD..89f3508M, 2011JCAP...07..025K, 2012PhRvL.109b1302P, 2012PhRvD..86b3518G}.
Moreover, in contrast
to observations of the Sunyaev-Zel'dovich (SZ) effect (or y distortions),
it is important to find the monopole part of the signal when measuring
$\mu$ distortions. This means  that the use of the difference in signals
from two different directions is not possible. Therefore,
the instrument should be well calibrated, and radiation emitted by the optics should be taken into account. This radiation is a barely modeled superposition of
radiations of different temperatures coming from different parts of an unevenly
cooled surface of the primary mirror, which can change during flight.

As a rule, the foreground spectra are described by analytical expressions
that depend on the parameters. The distribution of
parameters in the observed signal can, in principle, be arbitrary;
i.e., the exact shape of the foreground spectrum is hardly predictable.
A smart way of “rethinking" how to solve such a problem
was proposed in \citep{2017MNRAS.472.1195C}, where
a moment approach was introduced, but it extends the list of spectra to
be separated from the $\mu$ signal. Additionally, this approach implies strict
assumptions on the possible variation of the parameters.

The approach described here is completely different.
A method based on finding a special operator (“response”) applied to
the observed signal is proposed. This response minimizes the contribution
from foregrounds with parameters that are within a limited region of their
possible variations. The size and configuration of such a region can be
arbitrary and should be preestimated.
At the same time, the response to the normalized
$\mu$ signal itself in this algorithm is constant. It is shown below that,
when sufficient sensitivity is reached, the response to foregrounds
becomes negligibly small compared to the response to the $\mu$ signal.
Therefore, instead of modeling and disentangling the
foreground spectra with the necessary accuracy, the described algorithm
eliminates the contribution from any set of such foregrounds. It is important
to emphasize that this approach can be applied to any observation with a
poorly defined foreground radiation spectrum.

To briefly demonstrate the effectiveness of our approach, we restrict our
analysis to three foreground components: interstellar dust,
cosmic infrared background (CIB), and radiation from the telescope optics.
We use a modified blackbody to describe the emission from these three components
\citep{1984ApJ...285...89D}.  A simple modified blackbody may not be
suitable to approximate the interstellar medium dust SED at certain
sensitivity levels \citep{2021ApJ...914...68Z,2016ApJ...826..101K}.
However, the radiation from dust can be fitted with good accuracy by a
linear combination of modified blackbody spectra. For example, the
two-component dust model can accurately reproduce the emission observed
from dust in the diffuse interstellar medium of the Milky Way at 0.1-mm—3-mm
wavelengths \citep{1999ApJ...524..867F}. The rationalization of the choice
between alternative fitting methods, among other ideas, is
discussed in \citep{2015ApJ...814....9K}.

The  outline of this paper is as follows: In Sec. II the algorithm for
separation of $\mu$ distortion from foregrounds with poorly defined spectral
shapes is proposed.
Section III demonstrates the numerical results of applying the algorithm:
first, for the case with a single foreground and a single parameter, and
then for a more general case.
Brief conclusions are given in Sec. IV.
\vspace{-0.5cm}
\section{ Separation of the {\texorpdfstring{\Large $\mu$}{Lg}} signal from foregrounds with poorly defined spectral shapes}
In this section the algorithm for separation of the $\mu$-type distortion
from foregrounds is proposed. The signal
that we need to isolate from the total observed spectrum has the
following form \citep{2017MNRAS.471.1126A}:
\begin{equation}
    I_\mu=I_0\frac{x^4e^x}{(e^x-1)^2}\left(\frac{1}{b}-\frac{1}{x}\right)\mu,
\end{equation}
where $x=h\nu/kT_0$ and the CMB temperature is $T_0=2.72548$ $K$  \citep{1990ApJ...354L..37M,Fixsen_2009}.
The same estimated values for constants $b$, $I_0$, and $\mu$ as in
\citep{2017MNRAS.471.1126A} are used: $I_0=270$ $MJy/sr$, $\mu=2\times 10^{-8}$,
and $b=2.1923$. The total observed signal can be written as follows:
\begin{equation}
    S(\nu)=a_\mu I_\mu(\nu)+\sum\limits_{m=1}^MI_m(\nu),
\end{equation}
where $a_\mu$ is the amplitude to be found and $I_m(\nu)$ are $M$
different foregrounds of various origin.

To study the spectral properties of signals like $\mu$ distortion or the
Sunyaev-Zel'dovich effect, a device with a relatively low spectral resolution,
such as a Fourier-transform spectrometer (FTS), is usually used. It can measure the
spectrum from the minimum $\nu_{min}$ to the maximum $\nu_{max}$ frequency in multiple
frequency channels $\nu_j$, $j=1,..,J$, with the width of each channel
$\Delta\nu=\nu_{j+1}-\nu_j$. Thus, the discrete signal $S_j$ [or vector
${\bf S}=(S_1,...,S_J$)] that we measure is
\begin{equation}
  \begin{array}{l}
    \vspace{0.2cm}
    S_j=a_{\mu}I_{\mu}^j+\sum\limits_mI_m^j+N_j,\hspace{0.2cm}j=1,..,J\\
    I_{\mu}^j=\int\limits_{{}_{\nu_j-\frac{\Delta\nu}{2}}}^{{}^{\nu_j+\frac{\Delta\nu}{2}}}
    \hspace{-0.2cm}I_{\mu}(\nu)\frac{d\nu}{\Delta\nu},\hspace{0.3cm}
    I_m^j=\int\limits_{{}_{\nu_j-\frac{\Delta\nu}{2}}}^{{}^{\nu_j+\frac{\Delta\nu}{2}}}
    \hspace{-0.2cm}I_m(\nu)\frac{d\nu}{\Delta\nu},   
 \end{array}   
\end{equation}
where $N_j$ is the random noise for the $j$th frequency channel with zero mean
and variances $\langle N_iN_j\rangle=C_{ij}$. The covariance matrix of the noise
is expected to be close to the diagonal one: $C_{jj}=\sigma_j^2$
and $C_{ij}=0$ if $i\ne j$. The values of $\sigma_j$
depend on the photon noise coming from the sky and from the telescope
optics,  FTS frequency range $(\nu_{min}:\nu_{max})$, spectral resolution
$\Delta\nu$, number of FTS frequency bands, number of independent beams, and the
integrating time (duration of observations). 

In the general case, each $I_m$ depends on
$L$ parameters $p_\ell$, $\ell=1,..,L$, and each of the
observed foregrounds can be written as follows:
\begin{equation}
  \begin{array}{l}
    \vspace{0.2cm}
    I_m^j(\nu)=\int\limits_{\Omega} a_m({\bf P})
    f_m(\nu_j,{\bf P})d{\bf P},\\
    d{\bf P}=dp_{{}_1}dp_{{}_2}\cdot\cdot dp_{{}_L},
    \end{array}
\end{equation}
where ${\bf P}=(p_{{}_1},.,p_{{}_L})$ is the set of parameters, $f_m(\nu_j,{\bf P})$
are the functions representing the foreground spectra
(as a rule, described by an analytical formula), $\Omega$ is the parameter change region, and
$a_m$ are the amplitudes of the foreground radiation as functions of
parameters ${\bf P}$. Thus, if, for example, $a_m({\bf P})$ has the form of
a delta function $a_m({\bf P})=A_m\cdot\delta({\bf P}-{\bf P_m})$, then
the foreground spectrum
with index $m$ will have a template with well-defined parameters ${\bf P_m}$
and the amplitude $A_m$: $ I_m^j(\nu)=A_m\cdot f_m(\nu_j,{\bf P_m})$. Since we want to make our approach
as model independent as possible, we treat the functions $a_m({\bf P})$
as random with unknown properties. We impose very mild restrictions on
these functions as follows :\\
1. The integrated absolute values of the amplitudes $a_m$ should be less
than certain (preestimated) values $A_m$:
\begin{equation*}
  \begin{array}{l}
    \vspace{0.2cm}
    \int\limits_\Omega\mid a_m({\bf P})\mid d{\bf P}< A_m,\\
    a_m({\bf P})=0\hspace{0.3cm}for\hspace{0.3cm}{\bf P}\notin\Omega.
  \end{array}
\end{equation*}
\\
2. For foregrounds of different origins, random functions $a_m$ are independent
of each other, and consequently, $a_m$ and $a_k$ are uncorrelated if
$m\ne k$. This assumption is not exactly correct, and possible correlations can
be taken into account for a more detailed analysis.
\begin{center}
  \it{The algorithm}
\end{center}
The total observed signal ${\bf S}$ can be naturally divided into three parts
(three vectors):
\begin{equation}
  \begin{array}{l}
    \vspace{0.2cm}
   {\bf S}=a_\mu{\bf I_{\boldsymbol\mu}}+{\bf F}+{\bf {N}},\\
   {\bf F}=(F_1,..,F_J),\hspace{0.2cm}F_j=\sum\limits_mI_m^j,\\
   {\bf N}=(N_1,..,N_J)
   \end{array}
\end{equation}
where ${\bf I_{\boldsymbol\mu}}$ is the $\mu$ signal, ${\bf F}$ is the total
foreground,
and ${\bf N}$ represents the random noise.
The task of the algorithm is to find the  optimal vector of weights
$\boldsymbol\omega=(\omega_1,..,\omega_J)$
for frequency channels, which should have the following property:
\begin{equation}
  \boldsymbol\omega{\bf\cdot S^T}=\sum\limits_{j=1}^{J}\omega_jS_j
  \rightarrow a_\mu\hspace{0.2cm}
  for\hspace{0.2cm}\sigma_j\rightarrow 0,\hspace{0.2cm}j=1,..,J.
\end{equation}
Thus, the summation of the total observed signal over all channels with
appropriate weights should bring us as close as possible to the estimation of the
$\mu$ distortion amplitude $a_\mu$.

We call the scalar product $\boldsymbol\omega{\bf\cdot S^T}=R({\bf S})$
the response to the signal:
\begin{equation}
    R({\bf S})=a_\mu R({\bf I_{\boldsymbol\mu}})+R({\bf F})+R({\bf N}).
 \end{equation}
The first condition imposed on the weights is quite obvious:
\begin{equation}
R({\bf I_{\boldsymbol\mu}})=\sum\limits_j\omega_jI_\mu^j=1.
\end{equation}
The second
condition should minimize the response to the remaining part of the signal in
Eq. (7).

The mean square of the response to the foreground $R({\bf F})$ can be written as follows [see Eqs. (4) and (5)]:
\begin{equation}
    \langle R^2({\bf F})\rangle=
    \langle\sum\limits_{m=1}^Ma_m^2({\bf P})\left[\sum\limits_{j=1}^J
    f_m(\nu_j,{\bf P})
      \cdot\omega_j\right]^2\rangle. 
\end{equation}
According to our assumptions above about $a_m({\bf P})$, one can
write down the following inequality:
\begin{equation}
  \begin{array}{l}
    \vspace{0.2cm}
    \langle R^2({\bf F})\rangle<\sigma_{F,max}^2=
    \sum\limits_{i,j=1}^J\left[\sum\limits_{m=1}^MA_m^2q_{ij}^m\right]\omega_i\omega_j,\\
    q_{ij}^m=\frac{1}{V_\Omega}\int\limits_{\Omega}
    f_m(\nu_i,{\bf P})f_m(\nu_j,{\bf P})d{\bf P},
 \end{array}   
\end{equation}
where $V_\Omega$ is the volume of the $\Omega$ region. The
integrals $q_{ij}^m$ can be precalculated for all types of foreground
($m=1,...,M$) numerically or, in some particular cases, analytically depending
on the configuration of the parameter region  $\Omega$. 

Since $\langle R^2({\bf N})\rangle=\sum\limits_{i,j}C_{ij}\omega_i\omega_j$, the
minimization of the response to the foreground and to the noise is achieved with
weights $\omega_j$ corresponding to the minimum of the quadratic form $Q$:
\begin{equation}
  \begin{array}{l}
    \vspace{0.2cm}
 \langle\left(R({\bf F})+R({\bf N})\right)^2\rangle=
 \langle R^2({\bf F})\rangle+\langle R^2({\bf N})\rangle<Q,\\
 Q=\sum\limits_{i,j=1}^J\left[\sum\limits_{m=1}^MA_m^2q_{ij}^m+C_{ij}\right]
 \omega_i\omega_j.
 \end{array}
\end{equation}
Finally, one can find the coefficients $\omega_j$ for which
the minimum of the function $Q(\omega_1,..,\omega_{J})$ is reached:
\begin{equation}
  \begin{array}{l}
    \vspace{0.2cm}
    \frac{\partial Q}{\partial\omega_j}=0,\hspace{0.2cm}j=2,..,J,\\
    \omega_1=\frac{1}{I_\mu^1}-\sum\limits_{m=2}^J\omega_j\frac{I_\mu^j}{I_\mu^1}.
 \end{array}   
\end{equation}
Thus, $\omega_j$ calculated by Eq. (12) represent the optimal set of weights for
estimating the amplitude $a_\mu$. In fact, the solution of the Eq. (12) is
equivalent to the matched filter \citep{1999CQGra..16A.131S, 1999PhRvD..60b2002O, 2011LRR....14....5P,2021MNRAS.507.4852Z,2002MNRAS.336.1057H, 2017MNRAS.471.1167W, 2019MNRAS.489..401Z}
with covariance matrix ${\bf Q}=[Q_{ij}]$ and the template in the
form of the $\mu$ signal: 
\begin{equation}
  \begin{array}{l}
 \vspace{0.2cm}
 Q_{ij}=\sum\limits_{m=1}^MA_m^2q_{ij}^m+C_{ij},\\
 {\boldsymbol\omega}=\alpha\cdot{\bf Q}^{-1}{\bf I_\mu},
 \end{array}
\end{equation}
where the coefficient $\alpha$ is determined by the normalization in Eq. (8).
Note that instead of inverting the matrix ${\bf Q}$, it is much easier to solve the system of equations in Eq. (12). At low values of photon
noise (high sensitivity), the eigenvalues of this matrix can differ from
each other by many orders of
magnitude, which makes the process of inverting a large ${\bf Q}$ matrix
unstable.

To evaluate the efficiency of the algorithm,
it is convenient to use the following notations:
$\sigma^2_{{}_F}=\langle R^2({\bf F})\rangle$, $\sigma^2_{{}_N}=\langle R^2({\bf N})\rangle$.
The estimated amplitude $\tilde{a}_\mu$ coincides with the true amplitude $a_\mu$
with an accuracy of:
\begin{equation}
    \tilde{a}_\mu=a_\mu\pm\sqrt{\sigma^2_{{}_F}+\sigma^2_{{}_N}}.
\end{equation}
According to the notations in Eqs. (1) and (2), the expected amplitude in the considered
model is $a_\mu=1$. According to Eq. (10), $\sigma_{{}_{F,max}}>\sigma_{{}_F}$, and our estimate of the total variance is always overestimated:
$\sqrt{\sigma_{{}_{F,max}}^2+\sigma_{{}_N}^2}>\sqrt{\sigma_{{}_F}^2+\sigma_{{}_N}^2}$.

It should be noted that the choice of the two conditions indicated above
(on which the calculation of the matrix ${\bf Q}$ is based) cannot ensure
that the truly optimal coefficients are found. A more subtle approach would
be to restrict the functions $a_m({\bf P})$ from above in the following way:
\begin{equation}
    \mid a_m({\bf P})\mid<A_m({\bf P}),\hspace{1cm}{\bf P}\in \Omega.
\end{equation}
Nevertheless, the lack of information about the foregrounds forces us to
sacrifice the accuracy of the $\mu$ signal amplitude estimation. Otherwise,
the risk remains that an incorrect foreground model will lead to
misinterpretations of the observational data.  A more detailed foreground
modeling approach could, in principle, provide better coefficients
${\boldsymbol \omega}$, but this is outside the scope of our article.
It should also be noted that,
in reality,
$\langle R({\bf F})\rangle\ne 0$. This means that the $a_\mu$
estimate in our assumptions can be biased. Since we leave the distribution of
parameters unknown, we do not attempt to make any corrections for the bias.
Thus, the unknown bias is “hidden" in the total variance. In the next section, we give an example of a foreground model with a more or less realistic
distribution of parameters and show that this bias is small compared to the
variance.

\section{Extraction of {\texorpdfstring{\Large $\mu$}{Lg}}  distortion from a signal with foregrounds (numerical results)}

This section demonstrates the effectiveness of the algorithm in extracting the
$\mu$ signal from the observed spectrum in the presence of various foregrounds.

The contribution to the observed spectrum from some
of these foregrounds can be the sum of emissions with various uncertain
parameters. 

For clarity, let us start with the problem for a
single parameter and then proceed to demonstrate a more general case.

\subsection{Unknown combination of graybody spectra as an example
  of a foreground}

\begin{figure*}[!htbp]
  \includegraphics[width=0.49\textwidth]{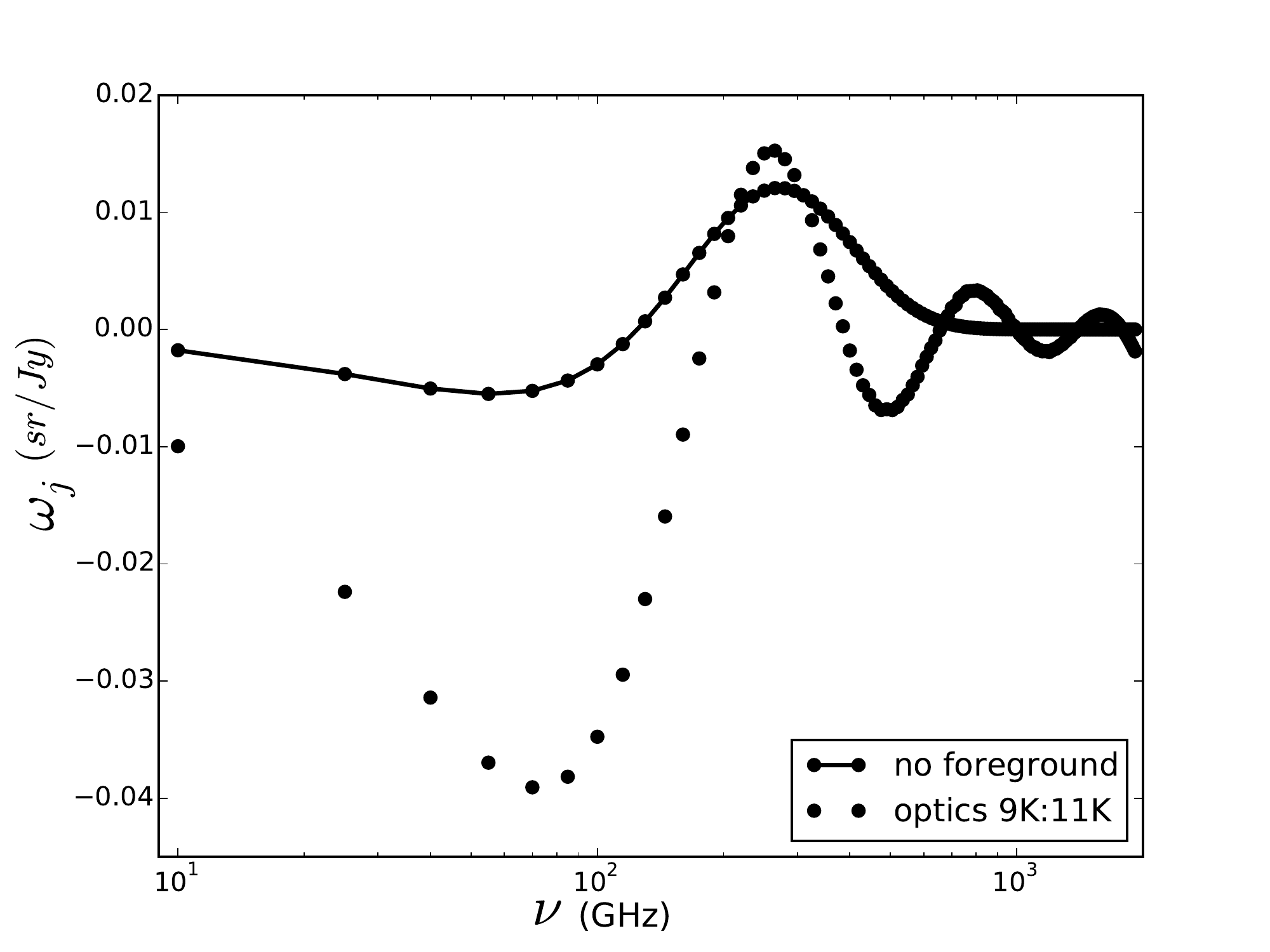}
  \includegraphics[width=0.49\textwidth]{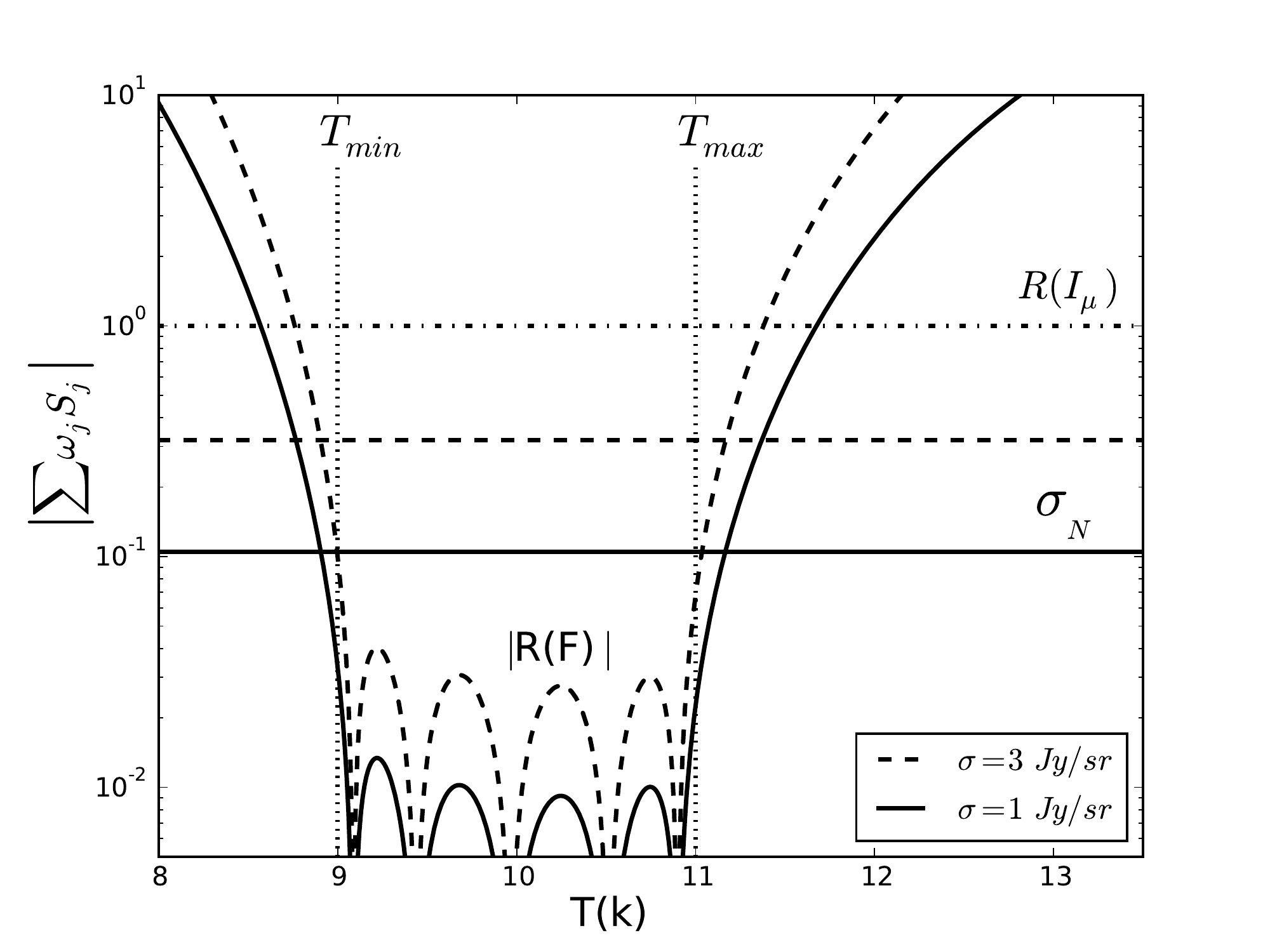}
  \caption{Results of the algorithm application when the foreground is 
    an unknown superposition of graybody spectra with
    temperatures distributed in any possible way between 9 K and 11 K.
    We assume emissivity
    $\int\limits_{9K}^{11K}\mid a(T)\mid dT<10^{-3}$.
    Left panel: optimal weights $\omega_j$ for $\sigma=3$ $Jy/sr$.
    The points connected by the solid line show $\omega_j$ when there is no
    foreground.
    Right panel: maximum possible
    absolute value of the response to the foreground $R({\bf F})$ as functions of
    temperature for $\sigma=3$ $Jy/sr$
    and $\sigma=1$ $Jy/sr$ shown in dashed and solid lines, correspondingly, assuming that all radiation
    is concentrated at one temperature $T$: $F(\nu)=10^{-3}\cdot B(\nu,T)$.    
    Any combination of sources with different temperatures distributed between
    9 K and 11 K with a restriction on $a(T)$ will give a response of less than
    $\frac{1}{(T_{max}-T_{min})}\int\limits_{T_{min}}^{T_{max}}\mid R({\bf F})\mid dT$. Horizontal dashed and solid lines represent
    the response to the noise. The horizontal dashed-dotted line is the response
    to the
  $\mu$ signal. Vertical lines limit the region of temperature variation.}
\end{figure*}

\begin{figure*}[!htbp]
  \includegraphics[width=0.49\textwidth]{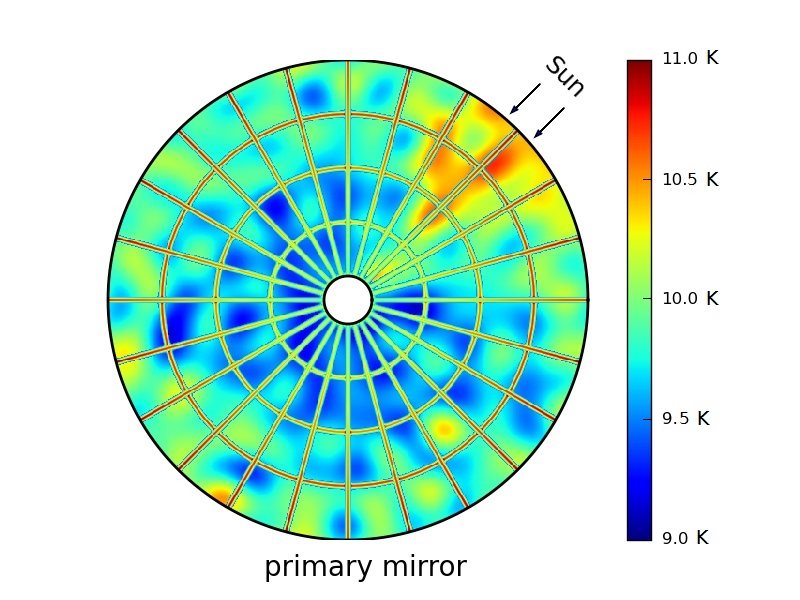}
  \includegraphics[width=0.49\textwidth]{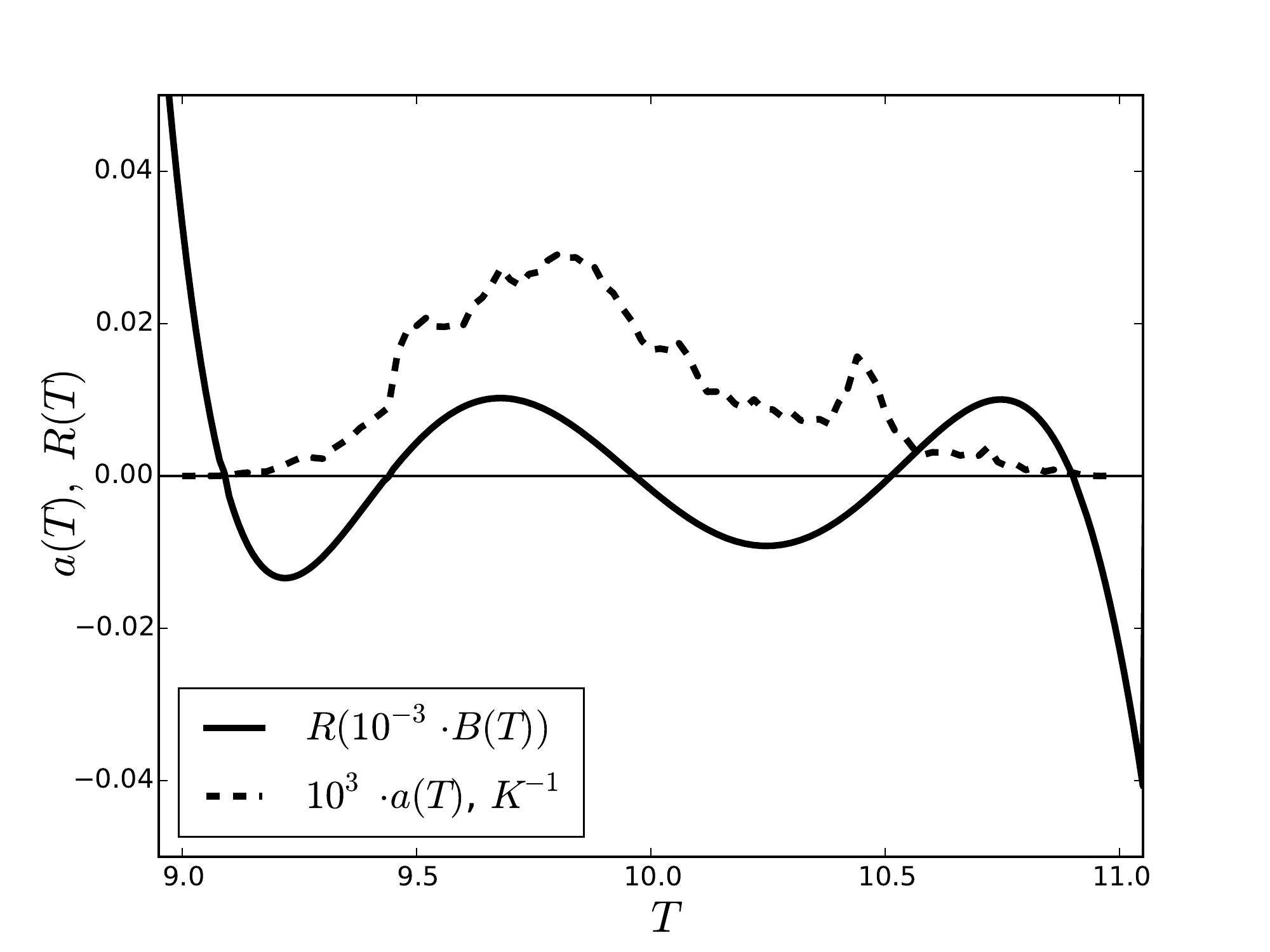}
  \caption{Simplified model of the foreground created by the telescope's
    primary mirror.
    Left panel: simulated temperature distribution over the surface of the mirror in the experiment \citep{2021PhyU...64..386N}. The gaps between the reflective panels have a slightly higher temperature than the panels themselves. Since the cooling machines are close to the center, the interior of the mirror is cooled more efficiently than the peripheral panels. The hot spot oriented at approximately 2 o'clock exists due to the corresponding orientation of the telescope relative to the Sun. This spot moves with time and makes a complete revolution around the mirror in one year. Right panel: amplitude distribution as a function of temperature $a(T)$ shown as a dashed line. The narrow peak at approximately 10.5 K corresponds to the contribution to the radiation from the gaps between the panels. The solid line shows the response to the graybody foreground when all radiation is concentrated at temperature $T$; i.e., $a(T)$ has the form of the delta function:
    $a(T')=10^{-3}\cdot\delta(T'-T)$ (same as in Fig. 1 for the photon noise
    $\sigma=1$  $Jy/sr$).} 
\end{figure*}

The simplest case is a problem with the foreground in the form of a superposition
of graybody spectra:
\begin{equation}
  \begin{array}{l}
    \vspace{0.2cm}
    I_{gb}(\nu)=\int\limits_{T_{min}}^{T_{max}} a(T)B(\nu,T)dT,\\
    B(\nu,T)=\frac{2(kT)^3}{(hc)^2}\frac{x^3}{e^x-1},
    \hspace{0.2cm}x=\frac{h\nu}{kT},
  \end{array}
\end{equation}
where $T_{min}:T_{max}$ is the range of possible temperature change from the
minimum to the maximum value. This range plays the role of the $\Omega$ region in the case
of a single parameter (temperature). One can always estimate
(for example, for a telescope's primary mirror) this range of temperature variations as well as the maximum possible value for
the mirror emissivity function:
$\int\limits_{T_{min}}^{T_{max}}\mid a(T)\mid dT<A_{max}$. The observed signal is
\begin{equation}
  \begin{array}{l}
    \vspace{0.2cm}
    S_j=a_\mu I_\mu^j+\int\limits_{T_{min}}^{T_{max}} a(T)B_j(T)dT+N_j,\\
    B_j(T)=\int\limits_{{}_{\nu_j-\frac{\Delta\nu}{2}}}^{{}^{\nu_j+\frac{\Delta\nu}{2}}}
    \hspace{-0.2cm}B(\nu,T)\frac{d\nu}{\Delta\nu},
  \end{array}
\end{equation}
For simplicity, we consider the covariance noise matrix to be a diagonal one.

In accordance with Eqs. (10) and (11), one can write an expression for the quadratic
form $Q$:
\begin{equation}
  \begin{array}{l}
    \vspace{0.2cm}
    Q=A_{max}^2\sum\limits_{i,j=1}^Jq_{ij}\omega_i\omega_j+
    \sum\limits_{j=1}^J\sigma_j^2\omega_j^2,\\
    q_{ij}=\frac{1}{T_{max}-T_{min}}\int\limits_{{}_{T_{min}}}^{{}_{T_{max}}}B_i(T)B_j(T)dT.\\  
 \end{array}   
\end{equation}
Thus, Eqs. (12) and (18) give us weights $\omega_j$.
If the amplitude of the noise greatly exceeds the possible contribution from
the foreground, then the optimal weights will be
$\omega_j\sim I_\mu^j/\sigma_j^2$ (as expected in the case of no foreground).
For the noise uniformly distributed over
all frequency channels ($\sigma_j=\sigma$), the weight function will have exactly the shape of the signal: $\omega_j\sim I_\mu^j$. By reducing the noise,
we begin to significantly change the optimal values of the weights and
thereby reduce not only the response to the noise $R({\bf N})$ but also the
response to an unknown foreground signal $R({\bf F}(T))$. The response to a  
foreground is a function of $T$, while the response to noise is just
a number.

In this numerical experiment the function $a(T)$ is random and unknown to us,
but
\begin{equation*}
  \int\limits_{9K}^{11K}\mid a(T)\mid dT<A_{max}=10^{-3}.
\end{equation*}  
The total number $J=128$ of
frequency channels $\nu_j$ were used from 10 GHz to 2 THz with the channel
width $\Delta\nu$=15 GHz.
Figure 1 demonstrates the maximum possible response to the foreground
$\mid R(10^{-3}\cdot{\bf B}(T))\mid>\mid R({\bf F}(T))\mid$ for two
different values of photon noise,
$\sigma$=3 Jy/sr and $\sigma$=1 Jy/sr. We can clearly see that for
sufficiently small $\sigma=\langle N_j^2\rangle$,
the optimally chosen coefficients $\omega_j$ provide a response to the
foreground that is negligible compared to the response to the signal,
$R({\bf I_{\boldsymbol\mu}})=1$.

Below we show an example of applying our algorithm to a real instrumental
foreground created by telescope optics. Figure 2 (left panel)
shows a simplified model
of the primary telescope mirror in the experiment \citep{2021PhyU...64..386N}.
This model is a 10-meter-diameter mirror cooled to 10 K and consisting of 96 panels.
Since the angular resolution is not a decisive factor in the study of
$\mu$ distortions, such a large mirror is not necessary. Nevertheless,
this experiment also involves the study of $y$ distortions and the
effects associated with the scattering of relic photons on plasma in galaxy
clusters (the SZ effect), where it is highly desirable to have a good resolution.
This picture shows the temperature distribution over the surface of an
unevenly cooled mirror. It is assumed that each surface element radiates as
a graybody with temperature $T$ and emissivity less than $10^{-3}$. The surface
temperature model of this mirror includes several terms:\\
$\bullet$ the average temperature T=10 K;\\
$\bullet$ the temperature gradient from the
center to the periphery (due to the internal panels
being cooled more efficiently);\\
$\bullet$ the hot spot due to one side of the telescope being heated
by the Sun;\\
$\bullet$ a random Gaussian temperature distribution with a characteristic scale of cold and
hot spots approximately corresponding to the size of the panels;\\
$\bullet$ the gaps between panels that are noticeably hotter than the rest of the surface.

The right panel of Fig. 2 shows the actual distribution of the
amplitude $a(T)$ over temperature along with the response to the
foreground when the amplitude is in the form of the delta function:
$a(T')=10^{-3}\cdot\delta(T'-T)$, $R({\bf F})=R(10^{-3}\cdot{\bf B}(T))$ (the
same as in Fig 1).
Thus, the response to the actual foreground created by the mirror is
\begin{equation}
    R({\bf F})=\int\limits_{9K}^{11K}a(T)R({\bf B}(T))dT.
\end{equation}

In this particular case, the response $R({\bf F})=0.091\sigma_{{}_{F,max}}$ is very
small compared to the estimated maximum possible variation.
As noted above, the average value of the response to the foreground is not
equal to zero. In order to find it we need to know the average distribution
$\langle a(T)\rangle$:
\begin{equation}
  \langle R({\bf F})\rangle=\int\limits_{9K}^{11K}\langle a(T)\rangle
  R({\bf B}(T))dT.
\end{equation}

In our particular model we can assume that this average distribution
does not differ much from the calculated $a(T)$ shown in Fig. 2. Thus, in real
parameter distributions the bias is not only less than
$\sigma_{{}_{F,max}}$
but, as a rule, it is significantly less than this overestimated variation.
Since in the general case we do not know the properties of the function
$a(T)$, we do not try to introduce any correction for the bias.

\begin{figure}
  \includegraphics[width=\columnwidth]{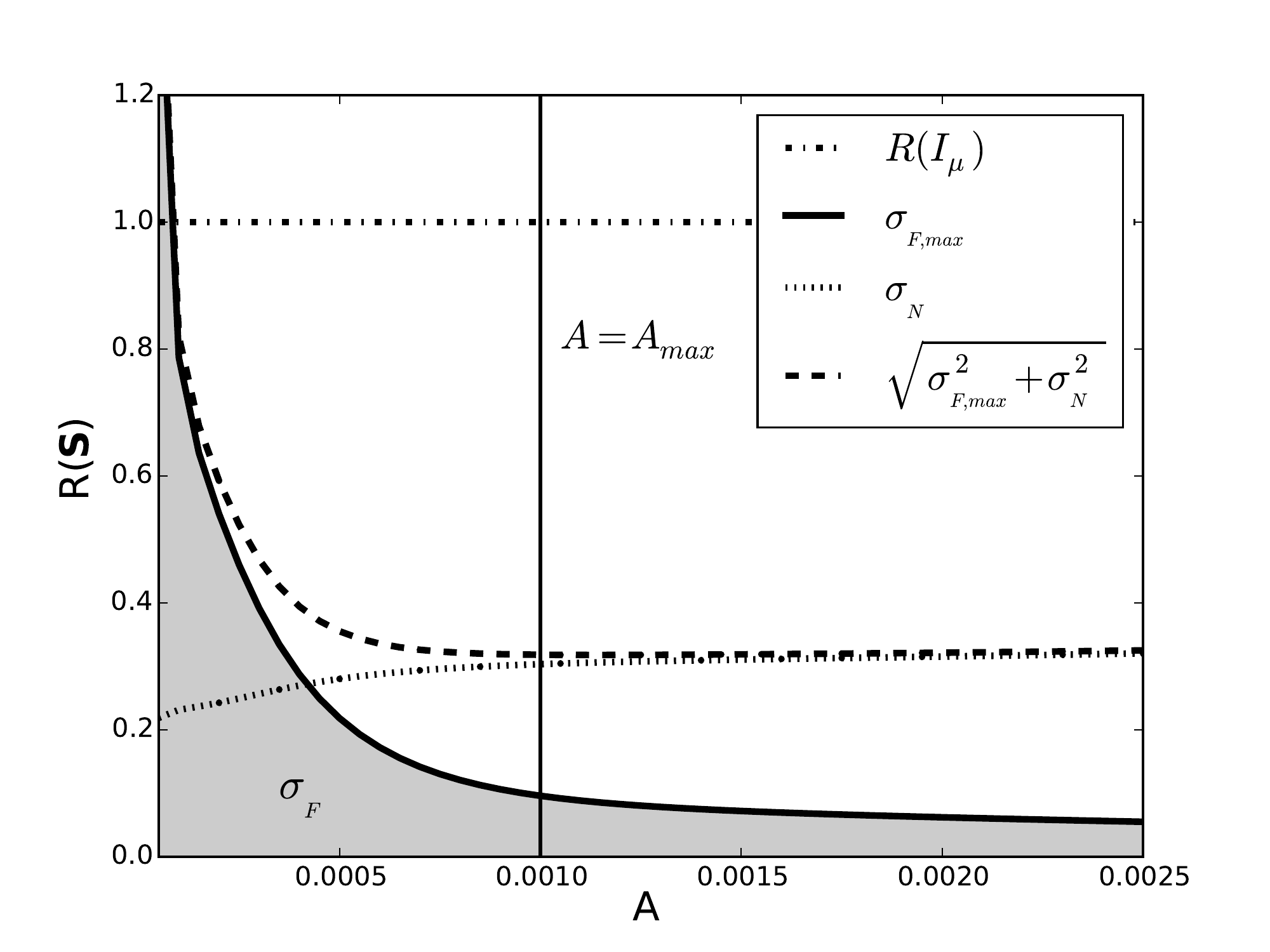}
  \caption{Dependence of $\sigma_{{}_F}$ and $\sigma_{{}_N}$ on the estimated
upper limit $A=\int\mid a(T)\mid dT$ of the amplitude.
Any combination of radiation sources in the form of a graybody with a temperature in the range from 9 K to 11 K and a total integrated amplitude less than $A$
will give a response $\mid R({\bf F})\mid$ that will be in the gray area below
the line $\sigma_{{}_{F,max}}$. The minimum of the total deviation
$\sqrt{\sigma_{{}_{F,max}}^2+\sigma_{{}_N}^2}$ is reached when the estimation of $A$ is correct: $A=A_{max}=10^{-3}$}
 \end{figure}

In this simplified example, it is easy to see that modeling the spectrum
emitted by the telescope optics is an extremely difficult (if not impossible)
task. Any attempt
to calculate such a spectrum (changing over the course of observations)
is complicated by a large number of factors that must be taken into account.
Our approach overcomes these difficulties. It is enough for us to know only
three quantities: the minimum and maximum possible temperatures of
the mirror surface, and its maximum possible emissivity. We also emphasize
that the optics radiation must be modeled by a combination of modified
blackbody radiation (the combination of graybody spectra is considered
here for simplicity).

Figure 3 demonstrates how important it is to correctly estimate the upper
limit of the amplitude $A_{max}$. It shows the dependence of
$\sigma_{{}_N}$ and $\sigma_{{}_{F,max}}$ on the estimate of the upper limit of
the amplitude $A$. Underestimation of this amplitude can lead to an
increase in the response to the foreground and an incorrect interpretation of
the data. At the same time, overestimation of this amplitude is not so
risky in this case. Nevertheless, in more general cases an overestimation
of the foreground amplitude can lead to a sharp increase in the response to
photon noise, which reduces the accuracy of $a_\mu$ estimation. The minimum
of the total deviation $\sqrt{\sigma_{{}_N}^2+\sigma_{{}_{F,max}}^2}$ of the response to
the signal from the true amplitude $a_\mu$ is reached when $A=A_{max}$. 
\begin{figure*}[!htbp]
  \includegraphics[width=0.325\textwidth]{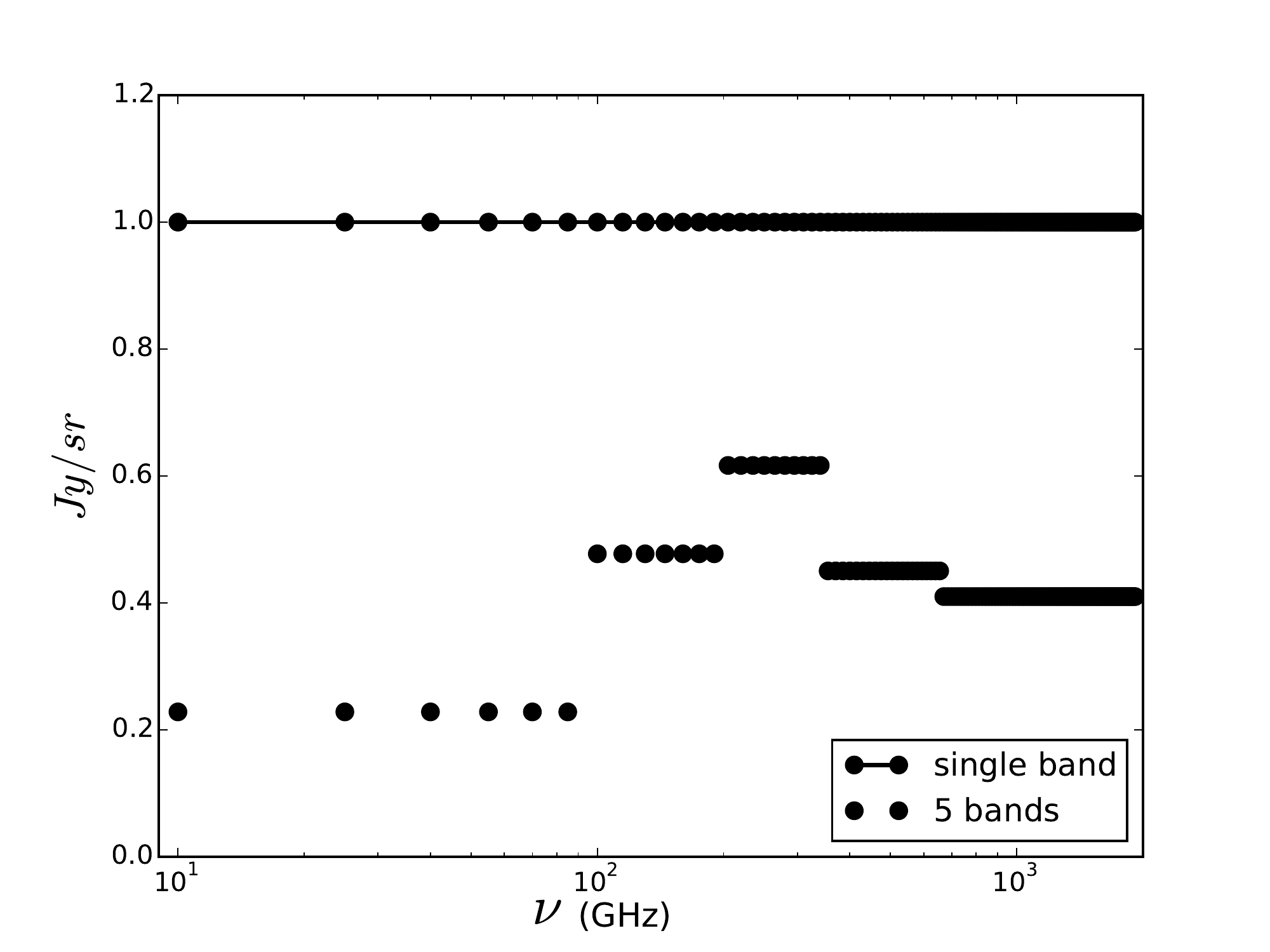}
  \includegraphics[width=0.325\textwidth]{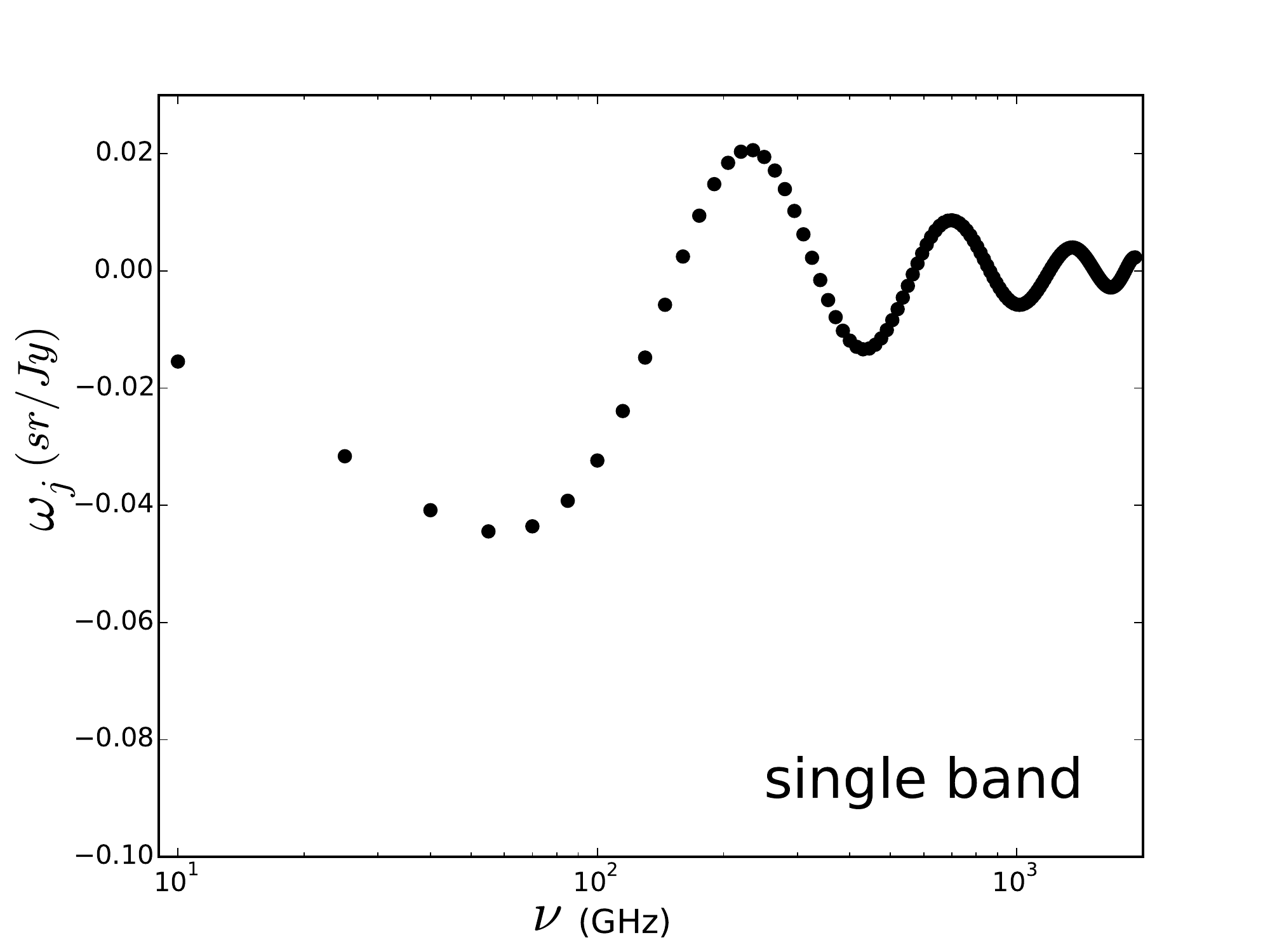}
  \includegraphics[width=0.325\textwidth]{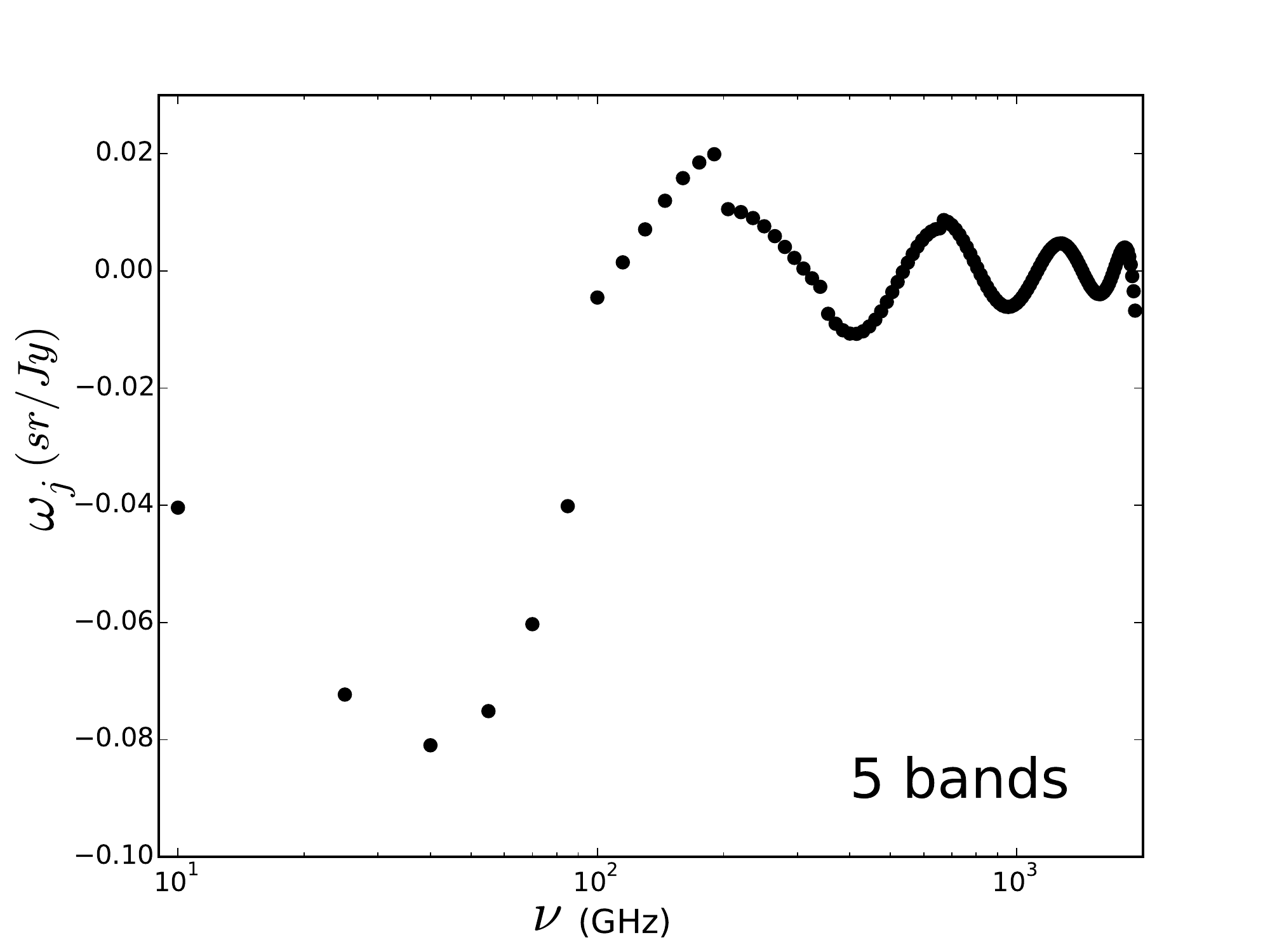}
  \includegraphics[width=0.325\textwidth]{aalg_res2_fig4d.pdf}
  \includegraphics[width=0.325\textwidth]{aalg_res2_fig4e.pdf}
  \includegraphics[width=0.325\textwidth]{aalg_res2_fig4f.pdf}  
  \caption{Separation of the $\mu$ signal from dust and CIB contamination.
    Top left: sensitivity for the frequency channels for one and
    five-band FTS.
    Bottom left: probability distribution function for parameters T and $\beta$ jointly for dust and infrared background.
    Top middle and top right: weights $\omega_j$
    for single-band and five-band sensitivity, correspondingly. 
    Bottom middle and bottom right:
    maximum possible foreground response
$\mid R({\bf F}(T,\beta))\mid$,
    $\int\limits_{\Omega}\mid a(T,\beta)\mid dTd\beta<A_{max}=10^{-6}$ for
    dust+CIB for single-band and five-band FTS, correspondingly.
    Black color indicates the area
    where the response to the foreground is
    greater than the response to the signal:
    $\mid R({\bf F}(T,\beta))\mid>R({\bf I_{\boldsymbol\mu}})=1$.
    Bright white filamentlike lines correspond to geometric points in the
    $T,\beta$ coordinate plane where $R({\bf F}(T,\beta))=0$. The responses
    to the noise are $\sigma_{{}_N}=0.124$ and $\sigma_{{}_N}=0.046$ for one
    and five bands, respectively.}
\end{figure*}

\subsection{Dust and CIB foregrounds}

As mentioned in the Introduction, dust and CIB contributions to the total
signal can both be written in the following form:
\begin{equation}
I_{{}_{dust,CIB}}(\nu,T,\beta)=\tau(\nu/\nu_0)^\beta B(\nu,T),
\end{equation}
where the reference frequency $\nu_0$ of 353 GHz is used.
Analogously to Eqs. (3) and (17), the total signal ${\bf S}=S_1,..,S_J$ is
\begin{equation}
  \begin{array}{l}
    \vspace{0.2cm}
    S_j=I_\mu^j+\int\limits_\Omega
    a(T,\beta)f(\nu_j,T,\beta)dTd\beta+N_j,\\
    f(\nu_j,T,\beta)=\int\limits_{{}_{\nu_j-\frac{\Delta\nu}{2}}}^{{}^{\nu_j+\frac{\Delta\nu}{2}}}
    \hspace{-0.2cm}(\nu/\nu_0)^\beta B(\nu,T)\frac{d\nu}{\Delta\nu}.
 \end{array}   
\end{equation}
In order to determine the boundaries of the parameter $(T,\beta)$ domain,
Planck data \citep{2014A&A...571A..11P,2016A&A...596A.109P} were used.
The probability distribution function for these parameters was calculated
using a 10-degree circular sky part centered at
$l=13.731^o$, $b=-73.946^o$; see Fig. 4 (bottom left panel).
The isocontour black lines limit the parameter region $\Omega(T,\beta)$.
(Note that dust and CIB areas can, in principle, overlap. This does not change
anything in our analysis since in this case we consider them as a single
foreground.) The probability of finding parameters outside this
region is less than 0.0002. At the same time, the maximum
allowable value of emissivity $\tau$ for the data we used does not
exceed $10^{-6}$:
$\int\limits_{\Omega}\mid a(T,\beta)\mid dTd\beta<A_{max}=10^{-6}$.
As in Sec. III A, 128 channels 
of 15 GHz width from 10 GHz to 2 THz were used. In order to compare the
effectiveness of different FTS configurations, the results for
two different cases are shown: single-band FTS and five-band FTS.
Both of their sensitivities (noise $\langle N_j^2\rangle$) are calculated
using \citep{1998IJIMW..19..931B,1986ApOpt..25..870L} for the same
integrating time (top left panel). Five-band FTS divides
the frequency range into five isolated parts. Therefore, it is not surprising
that it gives better sensitivity. The top middle panel and
top right panel show results of calculating
optimal weights $\omega_j$ for one and five bands, correspondingly. Unlike single-band weights, the weight function for five bands has discontinuities at points
equal to the minimum and maximum frequencies of each band. Results
for the maximum possible foreground response for these two cases are shown in
the bottom middle and bottom right panels. It is clear
that the five-band configuration provides us not only with a better noise
response but also with a smaller and safer foreground response.

\subsection{Other foregrounds}

In the previous subsection the algorithm was applied to the case with dust and
CIB. We now look at how other foregrounds can be included. First,
we should add the radiation created by the optics of the telescope since it is described
by the same modified blackbody formula and depends on the same parameters.
In this case one more region is added to the two regions in the
${\bf P}=(T,\beta)$ plane in Fig. 4. This region corresponds to variations in temperature and spectral slope for the
optical system.
Its size and configuration depend on the properties of the primary mirror:
average temperature, cooling system characteristics, the quality of surface
grinding, etc. The next foregrounds to be added are the spectral distortions
associated with the CMB radiation: CMB anisotropy (CMBA), SZ effect (y  distortions),
and its first relativistic correction \citep{1998ApJ...499....1C}. (The CMB monopole spectrum is well known
and can be subtracted from the total signal.)
The most “harmful" is the CMBA:
\begin{equation}
  \begin{array}{l}
    \vspace{0.2cm}
    I_{{}_{CMBA}}=\frac{2(kT_0)^3}{(hc)^2}\frac{x^4}{(e^x-1)^2}
    \cdot\frac{\Delta T}{T_0},\\
    x=h\nu/kT_0,
  \end{array}
\end{equation}
because its shape is exactly proportional to the first term in Eq. (1) for
$\mu$ distortion. This is not surprising because CMBA and $\mu$ distortion
have a similar physical origin. Therefore, particularly the second term in
Eq. (1) gives us an opportunity to measure chemical potential. This term manifests
itself mainly for $\nu<200$ GHz. Therefore, it is important to achieve good
sensitivity at relatively low frequencies. As for the maximum possible
CMBA amplitude, a safe estimate is
$\mid \frac{\Delta T}{T_0}\mid<A_{{}_{CMBA}}=10^{-4}$.
The shape of $I_{{}_{CMBA}}$ does not depend on any parameters ${\bf P}$,
but formally, we
consider this dependence to be a constant. Similarly, it is necessary to add
the SZ effect and the first relativistic correction to it. The upper limit
for their amplitudes depends on the specific position in the sky and the
presence of strong SZ sources.
Adding other foregrounds (synchrotron, free-free, etc.) with their floating
parameters one by one, we finally get a complete set of components that must be
taken into account when solving the problem of $\mu$ signal separation.
\section{Conclusions}
This paper presents a way to get rid
of cosmic foregrounds with poorly defined spectral characteristics
when measuring $\mu$ distortion.
The basis of this approach is the algorithm for finding special weights
for frequency channels. In the case of sufficient sensitivity, the sum
of the signal measurements with these weights (called the response) is weakly
sensitive to the presence of foregrounds with parameters lying in some
preestimated range of their possible variations. Therefore, the response
to the foregrounds becomes negligible in comparison with the response to
the $\mu$ signal. In this paper only
some types of foregrounds are considered. Applying the algorithm to all
possible foregrounds is the subject of a separate detailed
research.

It should be noted that this approach can be applied to experiments related to the study of phenomena associated with the SZ effect, for example Refs. \citep{2020PhRvD.101l3510N,2018PhRvD..98l3513E,2000MNRAS.312..159C,2016PhRvD..94b3513Y,2016JCAP...07..031S,2000ApJ...533..588I}, as well as to any physical
experiments with poorly defined foreground spectra. 

We would like to thank the referee for helpful comments
and a fruitful discussion.

This work is supported by Project No. 41-2020 of LPI's new scientific
groups and the Foundation for the Advancement of Theoretical Physics and Mathematics “BASIS,” Grant No. 19-1-1-46-1. 

\def\apj{Astrophys.~J}
\def\apjl{Astrophys.~J.,~Lett}
\def\apjs{Astrophys.~J.,~Supplement}
\def\an{Astron.~Nachr}     
\def\aap{Astron.~Astrophys}
\def\mnras{Mon.~Not.~R.~Astron.~Soc}
\def\pasp{Publ.~Astron.~Soc.~Pac}
\def\aaps{Astron.~and Astrophys.,~Suppl.~Ser}
\def\apss{Astrophys.~Space.~Sci}
\def\ibvs{Inf.~Bull.~Variable~Stars}
\def\japa{J.~Astrophys.~Astron}
\def\na{New~Astron}
\def\aspproc{Proc.~ASP~conf.~ser.}
\def\aspcs{ASP~Conf.~Ser}
\def\aj{Astron.~J}
\def\actaa{Acta Astron}
\def\araa{Ann.~Rev.~Astron.~Astrophys}
\def\caosp{Contrib.~Astron.~Obs.~Skalnat{\'e}~Pleso}
\def\pasj{Publ.~Astron.~Soc.~Jpn}
\def\memsai{Mem.~Soc.~Astron.~Ital}
\def\astl{Astron.~Letters}
\def\aipproc{Proc.~AIP~conf.~ser.}
\def\physrep{Physics Reports}
\def\jcap{J. Cosmol. Astropart. Phys.}
\def\baas{Bull. Am. Astron. Soc.}
\def\ssr{Space~Sci.~Rev.}
\def\azh{Astronomicheskii Zhurnal}

\bibliography{aalg_res2}



\end{document}